\documentclass{PoS}

\def\beq{\begin{equation}}
\def\eeq{\end{equation}}
\def\beqa{\begin{eqnarray}}
\def\eeqa{\end{eqnarray}}

\title{Soft-gluon resummations and NNNLO expansions}

\ShortTitle{Soft-gluon resummations and NNNLO expansions}

\author{\speaker{Nikolaos Kidonakis}%
         \thanks{This material is based upon work supported by the National Science Foundation under Grant No. PHY 1212472.}\\
        Kennesaw State University, USA\\
        E-mail: \email{nkidonak@kennesaw.edu}}

\abstract{I discuss the effectiveness of soft-gluon resummations in describing higher-order corrections. I present a comparison of recent resummation approaches and their relative successes in approximating complete NNLO corrections. I also discuss fixed-order expansions through NNNLO and present some recent applications to QCD hard-scattering processes, including top-antitop pair production, single-top production, and $W$ production at large transverse momentum.}

\FullConference{The European Physical Society Conference on High Energy Physics \\
                 18-24 July, 2013\\
                 Stockholm, Sweden}

\begin{document}

\section{Higher-order soft-gluon corrections}

Soft-gluon corrections are important in many QCD processes, such as the production of top-antitop pairs, single tops, $W$ bosons, jets, direct photons, etc.
The soft-gluon correction terms are of the form
$[\ln^k(s_4/M^2)/s_4]_+$ with $k \le 2n-1$ for the $n$th-order corrections, $M$ a hard scale, and $s_4$ the kinematical distance from partonic threshold. These corrections can be resummed via factorization and 
renormalization-group evolution (RGE) using the standard moment-space resummation formalism in pQCD. Complete results are now available at 
next-to-next-to-leading-logarithm (NNLL) accuracy, which involves the calculation two-loop soft anomalous dimensions \cite{NKPRL}.

Approximate next-to-next-to-leading-order (NNLO) and next-to-next-to-next-to-leading-order (NNNLO) master formulas for the soft-gluon corrections have been derived from the expansion \cite{NKNNNLO} of the resummed cross sections. The resummed calculations and their NNLO and NNNLO expansions are at the double-differential cross-section level.

Resummation follows from factorization properties of the 
cross section, performed in moment space, with $N$ the moment variable conjugate to $s_4$:  $\sigma=(\prod \psi) \; H_{IL} \, S_{LI} \; (\prod J)$, with  
$H$ the hard function, $S$ the soft-gluon function, and $\psi$ and $J$ functions for incoming and outgoing partons. The soft-gluon function describes noncollinear soft gluon emission and is dependent on the color structure of the partonic process. We use RGE to evolve the soft-gluon function  
$$
\left(\mu {\partial \over \partial \mu}
+\beta(g_s){\partial \over \partial g_s}\right)\,S_{LI}
=-(\Gamma^\dagger_S)_{LB}S_{BI}-S_{LA}(\Gamma_S)_{AI}
$$
where 
$\Gamma_S$ is the soft anomalous dimension - a matrix in 
color space and a function of the standard kinematical invariants $s$, $t$, $u$.
We determine $\Gamma_S$ from ultraviolet poles in dimensionally regularized 
eikonal diagrams.  $\Gamma_S$ is process-dependent and is calculated at two-loop accuracy.

The other functions in the refactorized cross section can be similarly evolved and the resummed partonic cross section in moment space takes the form  
\beqa
{\hat{\sigma}}^{res}(N) &=&
\exp\left[ \sum_i E_i(N_i)\right] \, \exp\left[ \sum_j E'_j(N')\right]\;
\exp \left[\sum_{i=1,2} 2 \int_{\mu_F}^{\sqrt{s}} \frac{d\mu}{\mu}\;
\gamma_{i/i}\left({\tilde N}_i, \alpha_s(\mu)\right)\right] \;
\nonumber\\ && \hspace{-20mm} \times \,
{\rm tr} \left\{H\left(\alpha_s\right)
\exp \left[\int_{\sqrt{s}}^{{\sqrt{s}}/{\tilde N'}}
\frac{d\mu}{\mu} \;
\Gamma_S^{\dagger}\left(\alpha_s(\mu)\right)\right] \;
S \left(\alpha_s\left(\frac{\sqrt{s}}{\tilde N'}\right)
\right) \;
\exp \left[\int_{\sqrt{s}}^{{\sqrt{s}}/{\tilde N'}}
\frac{d\mu}{\mu}\; \Gamma_S
\left(\alpha_s(\mu)\right)\right] \right\}
\nonumber 
\nonumber 
\eeqa
where the first two exponents are universal factors for collinear and soft emission from incoming and outgoing partons, the third exponent controls the factorization scale dependence, and the last two exponents involve the soft anomalous dimension matrices. 
The above equation resums $\ln^k N$ - we can expand it to fixed order 
and invert to momentum space to get $\ln^k(s_4/M^2)/s_4$.

\section{Relevance of threshold approximations}

\begin{figure}
\begin{center}
\includegraphics[width=9cm]{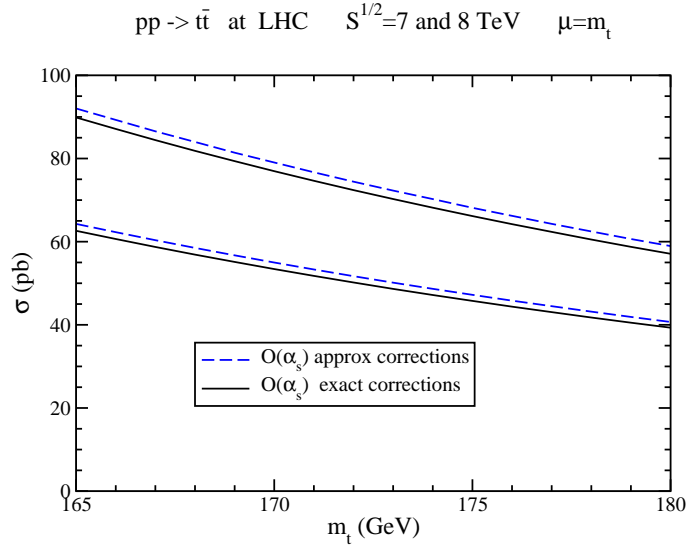}
\caption{Comparison of exact and approximate {\cal O}($\alpha_s$) corrections to the $t{\bar t}$ cross section at the LHC with 7 TeV (lower lines) and 8 TeV (upper lines) energy.}
\label{top1a1elhcplot}
\end{center}
\end{figure}

The threshold approximation works very well for LHC and Tevatron energies
for many processes. In particular it is an excellent approximation for top-pair production at the LHC: there is only 
$\sim$1\% difference between NLO approximate and exact total cross sections (see e.g. Fig. 1), and also differential distributions in transverse momentum, $p_T$, and rapidity \cite{NKtop}. This is also known to be true at NNLO for total cross sections, i.e. exact and approximate results are nearly identical.

There are several approaches to threshold resummation in the literature and they are reviewed in some detail in \cite{NKBP}. Major differences include the formalism used (moment-space pQCD vs SCET); and whether the resummation is for total-only or double-differential cross sections. 
Resummations for total-only cross sections use the threshold variable $\beta = \sqrt{1-4m_t^2/s}$ which vanishes at production (absolute) threshold, with zero top-quark velocities. Resummations for double-differential cross sections, $d\sigma/dp_T dy$, in single-particle-inclusive (1PI) kinematics use the variable $s_4= s+t_1+u_1$ which vanishes at partonic threshold (top quarks not necessarily at rest). For pair-invariant-mass (PIM) kinematics, $d\sigma/dM_{t\bar t}d\theta$, the variable is $1-z = 1-M_{t\bar t}^2/s$.
The more general approach is of course double-differential since it allows the calculation of $p_T$ or rapidity or other distributions, and partonic threshold is a more general definition of threshold.

\begin{figure}
\begin{center}
\includegraphics[width=10cm]{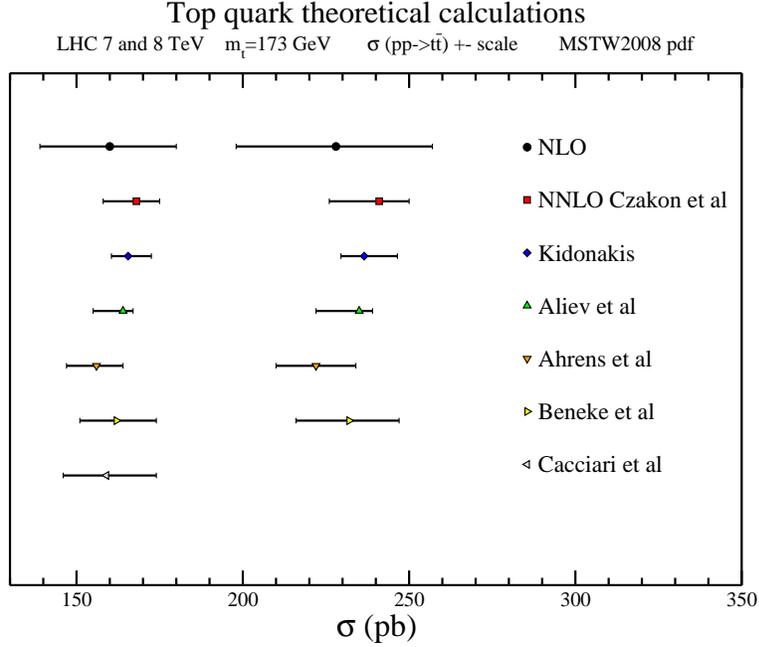}
\caption{Comparison of theoretical cross sections for $t{\bar t}$ production at the LHC with 7 TeV (bars on the left) and 8 TeV (bars on the right) energy.}
\label{theory14lhcplot}
\end{center}
\end{figure}

A comparison of various NNLO approximate approaches to exact NLO and NNLO results for $t{\bar t}$ production  at 7 TeV and 8 TeV LHC energy, all with the same choice of parameters, is shown in Fig. \ref{theory14lhcplot}. The result from the formalism used in this work \cite{NKtop} is very close to the exact NNLO and provides the best approximation: 
both the central values and the scale uncertainty are nearly the same as the exact NNLO.  The approximation used in our formalism \cite{NKtop} is also excellent for all collider energies and top quark masses under consideration. This agreement was expected from earlier comparisons of exact and approximate corrections at NLO, and analytical/numerical studies of NNLO corrections in 1PI and PIM kinematics.
Results on approximate NNNLO have already appeared in \cite{NKNNNLO} 
and more work on NNNLO is in progress.

The reliability of the theoretical soft-gluon results is very important since it provides confidence in applications to other processes, 
and the high quality of the approximation indicates that we have near-exact NNLO $p_T$ and rapidity distributions.

\section{Top quark production}

We present some results for top quark production, both in pair production and single-top production channels. We begin with top-quark differential distributions in top-antitop pair production. We use MSTW2008 NNLO pdf \cite{MSTW} in our numerical results.

\begin{figure}
\begin{center}
\includegraphics[width=9cm]{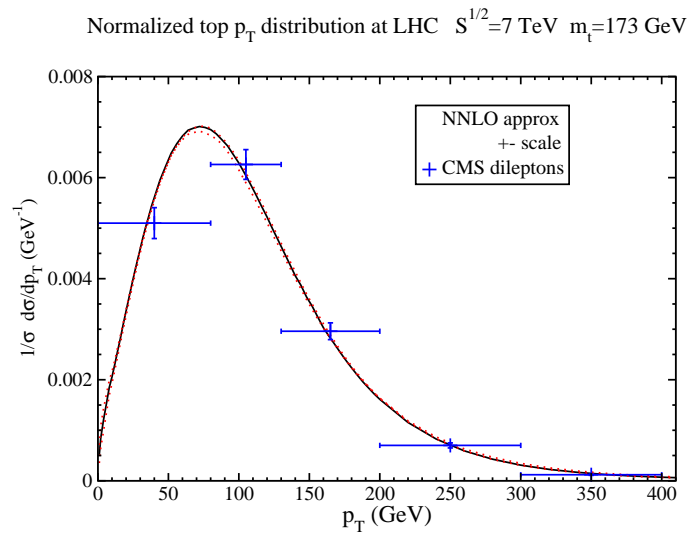} 
\caption{The top quark normalized $p_T$ distribution at approximate NNLO compared with CMS data in the dilepton channel at 7 TeV LHC energy.}
\label{toppt}
\end{center}
\end{figure}

Figure \ref{toppt} shows the normalized approximate NNLO top quark $p_T$ distribution at the LHC at 7 TeV energy. The central result is with scale choice $\mu=m_t$ and the variation with scale $m_t/2 < \mu < 2m_t$ is also shown. Recent results from CMS \cite{CMS7pty} in the dilepton channel are also displayed. The agreement between theory and CMS data is remarkable; in fact, as discussed in \cite{CMS7pty} the NNLO approximate result describes the data at both low and high $p_T$ significantly better than NLO predictions from event generators. This is also true for the CMS results in the $\ell$+jets channel \cite{CMS7pty}. Finally, there are more recent CMS results at 8 TeV LHC energy \cite{CMS8pty} and the same pattern holds.

The approximate NNLO top quark rapidity distribution has also been calculated.
Again, the agreement with recent results from CMS \cite{CMS7pty} in the $\ell$+jets and dilepton channels at 7 TeV LHC energy is excellent, and this is also true for the more recent CMS results at 8 TeV energy \cite{CMS8pty}.

\begin{figure}
\begin{center}
\includegraphics[width=10cm]{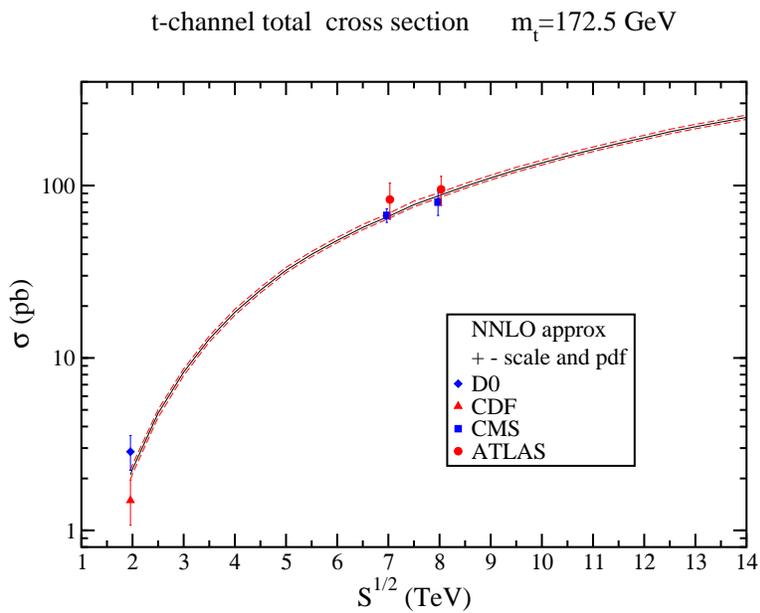}
\caption{The total cross section for $t$-channel single top plus single antitop production compared with Tevatron and LHC data.}
\label{tch}
\end{center}
\end{figure}

We continue with single-top quark production \cite{NKtch} and show some results for $t$-channel production.  Figure \ref{tch} shows the total NNLO approximate $t$-channel cross section as a function of collider energy, together with scale and pdf uncertainties. The agreement with Tevatron \cite{CDFts,D0ts} and LHC \cite{ATLAStch,CMStch} data is very good. 

\begin{figure}
\begin{center}
\includegraphics[width=9cm]{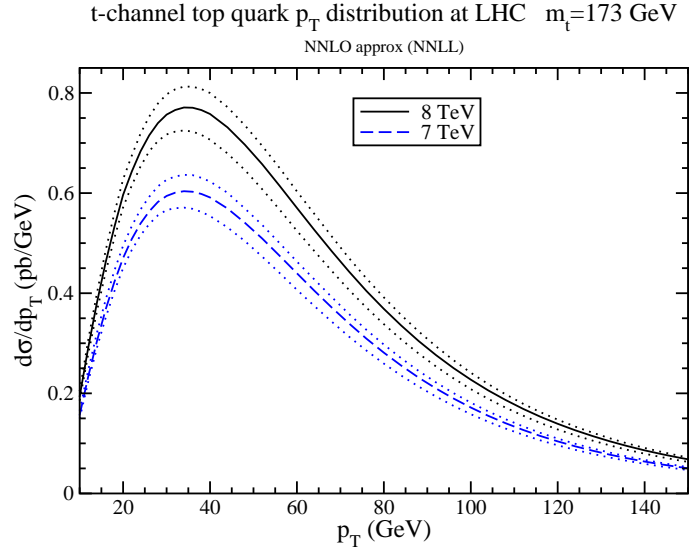}
\caption{The top-quark $p_T$ distribution in $t$-channel single-top production at the LHC.}
\label{tchpt}
\end{center}
\end{figure}

In Fig. \ref{tchpt} we show the top-quark $p_T$ distributions in $t$-channel single-top production at 7 and 8 TeV LHC energy, including the theoretical uncertainties from scale variation.

\section{$W$ production at large $p_T$}

Threshold resummation has also been successfully applied to $W$ production at large $p_T$ \cite{NKRG}.

\begin{figure}
\begin{center}
\includegraphics[width=9cm]{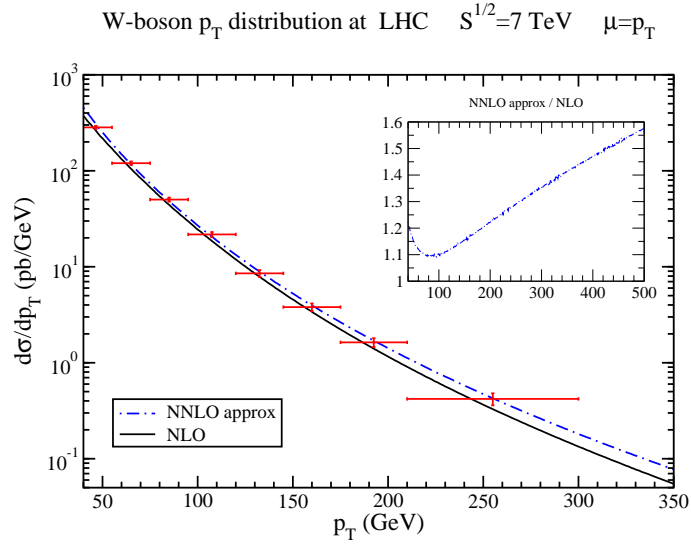}
\caption{$W$-boson NLO and approximate NNLO $p_T$ distributions at the LHC.}
\label{W7}
\end{center}
\end{figure}

\begin{figure}
\begin{center}
\includegraphics[width=9cm]{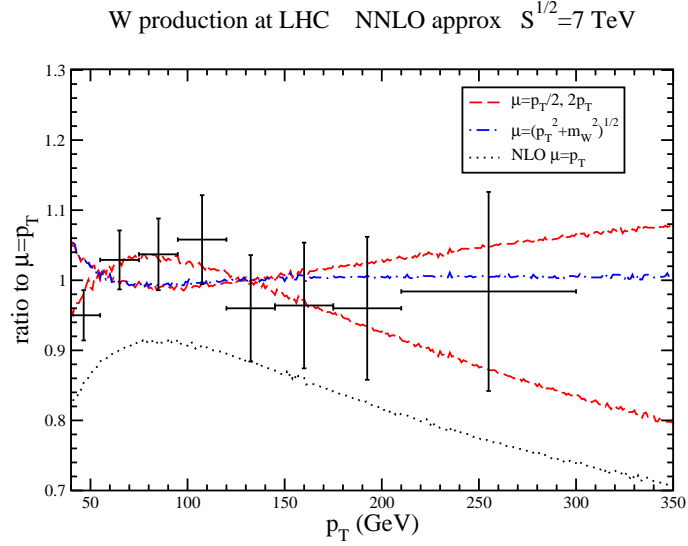}
\caption{Comparison of $W$-boson $p_T$ distributions with different scale choices.}
\label{W7mu}
\end{center}
\end{figure}

Figure \ref{W7} shows NLO and approximate NNLO results for the $W$-boson $p_T$ distribution at 7 TeV LHC energy, and the NNLO/NLO ratio is shown in the inset plot.  ATLAS data \cite{ATLAS-W} are also shown on the plot.
Figure \ref{W7mu} shows the ratio with various scale choices to the central result with $\mu=p_T$. The comparison to ATLAS data \cite{ATLAS-W} shows very good agreement between theory and experiment when NNLO corrections are included.

\end{document}